\documentclass[review,12pt,authoryear]{elsarticle}

\usepackage[enable,chapter]{easy-todo} 

\usepackage{enumitem}
\usepackage{graphicx}
\usepackage{amsmath}
\usepackage[linesnumbered,lined,commentsnumbered]{algorithm2e}
\usepackage[table]{xcolor}
\usepackage{amssymb}
\usepackage{geometry}
\usepackage{multirow}

\newdefinition{definition}{Definition}
\newtheorem{theorem}{Theorem}

\def\url#1{\expandafter\string\csname #1\endcsname}

\journal{Applied Soft Computing}

\begin{document}

\begin{frontmatter}

\title{Sparsifying Parity-Check Matrices \tnoteref{t1}}

  \author{Lu\'is~M.~S.~Russo \corref{cor1}}
\ead{luis.russo@tecnico.ulisboa.pt}
\address{INESC-ID and the Department of Computer Science and Engineering,\\
  Instituto Superior T\'ecnico, Universidade de Lisboa, Portugal}

\author{Tobias~Dietz \corref{c2}}
\ead{dietz@mathematik.uni-kl.de}
\address{ Department of Mathematics, Technische Universit\"at\\
  Kaiserslautern, Germany}

\author{Jos\'e~Rui~Figueira}
\ead{figueira@tecnico.ulisboa.pt}
\address{ CEG-IST, Instituto Superior T\'ecnico, Universidade de Lisboa, Portugal}

\author{Alexandre~P.~Francisco}
\ead{aplf@tecnico.ulisboa.pt}
\address{INESC-ID and the Department of Computer Science and Engineering,\\
  Instituto Superior T\'ecnico, Universidade de Lisboa, Portugal
}

\author{Stefan~Ruzika}
\ead{ruzika@mathematik.uni-kl.de}
\address{Department of Mathematics, Technische Universit\"at\\
  Kaiserslautern, Germany}

\cortext[cor1]{Corresponding author}

\begin{abstract}
  Parity check matrices (PCMs) are used to define linear error correcting codes and ensure reliable information transmission over noisy channels. The set of codewords of such a code is the null space of this binary matrix.
  We consider the problem of minimizing the number of one-entries in parity-check matrices. In the maximum-likelihood (ML) decoding method, the number of ones in PCMs is directly related to the time required to decode messages. We propose a simple matrix row manipulation heuristic which alters the PCM, but not the code itself. We apply simulated annealing and greedy local searches
  to obtain PCMs with a small number of one entries quickly, i.e.~in a couple of minutes or hours when using mainstream hardware. The resulting matrices provide faster ML decoding
  procedures, especially for large codes.
\end{abstract}


\end{frontmatter}

\section{Introduction} 
\label{sec:introduction}
In today's world, a fast and reliable wireless internet connection is
essential. During data transmission, the data may become perturbed due to
weather conditions, obstacles, or other data traffic. To achieve robustness, usually all data is encoded before the transmission and decoded after reception. While encoding is an easy task once a suitable code is chosen, decoding may be extremely costly in terms of time and it may produce errors. These two problems can for example occur when streaming of a video: The video may need much time to load and buffer or the video may be of bad quality.

The optimal way to decode a message is the so-called Maximum-Likelihood~(ML) decoding (see \citet*{helmling2012} for an introduction). It is well-known that ML decoding is NP-hard (\citet*{berlekamp}). In this paper, we aim to reduce the ML decoding time by altering the representation of the given code. In detail, we alter the code's underlying PCM such that the number of ones is reduced without changing the code itself. This can be done by adding one or more rows to another row which does not change the kernel of the matrix and thus does not change the code. \citet*{DBLP:conf/ict/GensheimerDRKW18} show that ML decoding works faster on sparse matrices. The authors also show that the number of one-entries of a PCM can be minimized by solving an integer program for every row of the matrix. Although this optimization has to be done only once for
each code, the computation time increases rapidly with increasing matrix size. Since each integer problem considered by \citet*{DBLP:conf/ict/GensheimerDRKW18} consists of $\mathcal{O}(n)$~variables and constraints, where $n$ is the number of columns, the authors did not compute an optimal solution for large matrices. Instead, approximations are
given. In this paper, we focus on obtaining fast algorithms that approximate an optimal matrix by applying simulated annealing.

The remainder of this article is organized as follows. Section~2 presents basic concepts, definitions, and notation. Section~3 is devoted to matrix transitions and the problem is formulated in the context of simulated annealing meta-heuristics with the aim of modifying the current PCM quickly. Section~4 contains computational results which show that our approach efficiently yields very good
approximations of the optimal sparse PCM in reasonable time frames, from a few minutes to a couple of hours, with commodity hardware (in some cases even greedy local searches obtain good results). Moreover, we also provide practical guidelines on how to select temperatures and cooling schedules for annealing. Section~5 presents other approaches used to deal with the same problem, namely those based in binary linear programming and solved by
powerful linear programming solvers. Finally, Section~6, presents the main conclusions and provides future research ideas.

\section{Coding Theory Basics} 
\label{sec:concepts}
\noindent This section introduces basic notations, a fundamental result, and an illustrative example.

\begin{definition}[Binary Linear Code]
  A \emph{binary linear code}~$C\subseteq\{0,1\}^n$ of size~$n$ is a linear subspace of $\{0,1\}^n$. The elements of $C$ are called \emph{codewords}.
\end{definition}

In particular, $0^n\in C$ and, for all $y,y'\in C$, it holds that $y+y'\in C$ where the addition is performed modulo~2.

\begin{definition}[Parity-Check Matrix]
\label{def:code}
  A binary linear code~$C$ can be represented by a \emph{parity-check
    matrix (PCM)} $H\in \{0,1\}^{(n-k)\times n}$ where
  $C=\left\{y\in \{0,1\}^n: Hy\equiv 0\mod 2\right\}$. If the rows of $H$
  are linearly independent, $k$ is the \emph{dimension} of $C$.
\end{definition}
Thus, a binary linear code can be seen as the kernel of a binary
matrix. With a parity-check matrix~$H$ given, it can be easily decided if a
given word~$y\in \{0,1\}^n$ belongs to the code~$C$.

\begin{theorem}[see \citep*{DBLP:conf/ict/GensheimerDRKW18}]
  Let $H\in\{0,1\}^{(n-k)\times n}$ be a parity-check matrix of a code~$C$ of dimension~$k$. Then $H'\in\{0,1\}^{(n-k)\times n}$ is a parity-check matrix of $C$ if and only if all rows of $H$ are elements of the span of the rows of $H'$.
\end{theorem}
\noindent
In particular, the code does not change if one or more rows are added to another row.

\noindent
The following matrix is a PCM~$H$ of a BCH code\footnote{In particular $n = 15$ is the codeword size and $k=7$ is the
  dimension.},
\begin{equation}
  \label{eq:1}
  H = \left(
  \begin{array}{cccccc>{\columncolor{black!20}}cccc>{\columncolor{black!20}}cc>{\columncolor{black!20}}c>{\columncolor{black!20}}c>{\columncolor{black!20}}c}
  1&0&1&1&1&0&0&1&1&0&0&0&0&0&0\\
  0&1&1&0&1&0&0&0&1&0&0&0&0&0&0\\
  0&0&1&1&0&1&0&0&0&1&0&0&0&0&0\\
  0&0&0&1&1&0&1&0&0&0&1&0&0&0&0\\
  0&0&0&0&1&1&0&1&0&0&0&1&0&0&0\\
  0&0&0&0&0&1&1&0&1&0&0&0&1&0&0\\
  0&0&0&0&0&0&1&1&0&1&0&0&0&1&0\\
  0&0&0&0&0&0&0&1&1&0&1&0&0&0&1\\
  \end{array}
  \right).
\end{equation}

The codewords are the vectors $y$ for which $H \cdot y \equiv \vec{0} \mod 2$,
where $\vec{0}$ is a vector of zeros. Therefore, the word $000000000000000$
is a codeword, likewise $000000100010111$ is also a codeword, since
$H \cdot y = (0 0 0 0 0 0 0 0)^T$, when $y$ is the vector representation
of this word. Note that for latter, $H \cdot y$ corresponds to
summing (modulo $2$) the columns highlighted in gray. This matrix contains $6$~ones in the first row and $4$~ones in each of the following $7$~rows, yielding a total of $34$~ones.

From basic linear algebra, it follows that a PCM can be modified in such a way that the underlying code is not changed. More precisely, given any invertible matrix~$S$, the matrix~$H' = S\cdot H$
yields an alternative PCM for the same code. For example, the first row of $H$ can be replaced by adding the second row to it. This  yields $H'$ where the first row is $(1 1 0 1 0 0 0 1 0 0 0 0 0 0 0)$ while the remaining rows are identical to those in $H$. This first row contains $4$~ones and, therefore, $H'$ contains $32$~ones in total.

Given some code, our goal is to compute a PCM~$H'$ with the minimal number of one entries. In the example above, $H'$ is sparser than
$H$. In fact, for this particular code, it can be proven that the minimal number of ones of any PCM equals $32$ and, therefore, $H'$ is one of the sparsest PCMs.

\section{Constructing Alternative Parity-Check Matrices} 
\label{sec:my-idea}
There may be several natural approaches to the problem of finding  an invertible matrix~$S$ such that $H'= S \cdot H$ is sparse\footnote{
Note that, for any word~$y$, we want to have
$H y \equiv \vec{0}$ iff $H' y \equiv \vec{0}$. The forward implication is
straight forward, if $H y \equiv \vec{0}$ then
$H' y \equiv S \cdot H y \equiv S \vec{0} = \vec{0}$. The implication in the reverse direction can be proven by assuming that
$H' y \equiv \vec{0}$, and because $S$ is invertible, we can apply
$S^{-1}$ to both sides of the equation and obtain that $H y \equiv \vec{0}$.}.

The approach proposed in this article consists of selecting an origin row~$i$ and a destination row~$j$ and adding row~$i$ to row~$j$. The matrix~$S$ representing this kind of elementary row operation is the unit matrix with an additional~$1$ in the entry~$(j,i)$. This process is repeated several times, until the resulting matrix is sufficiently close to the optimal value.

\subsection{Row Selection}\label{sec: ChooseRows}
A successful realization of this strategy entails several open issues, namely how to efficiently select the rows $i$ and $j$.

In particular, adding one row to another may sometimes increase the number of ones in $H^\prime$, although this process seems to be contradicting the objective. However, it is sometimes necessary to escape from local minima in the search space. We show that the simulated annealing meta-heuristic provides a good policy to guide the search process for a sparse parity-check matrix.

The main steps of the simulated annealing algorithm take into account the following aspects:
\begin{enumerate}
\item In the context of simulated annealing, we use $E(H^\prime)$ to represent the number of ones in $H'$, i.e., the energy of the current state.
\item We associate with each transition (i.e. adding one row to another) a probability~$e^{-d/T}$, where $d$
  is the variation of $E(H^\prime)$ and $T$ is the current temperature of
  the process.
\item If the transition maintains the value $E(H')$ constant then $d=0$ and
  this probability is $1$, in which case the transition is
  accepted. Likewise, if $d<0$, then the probability formula yields a value greater than
  $1$ and the transition is also accepted.
\item If $d>0$, then the transition represents an uphill movement and is not
  always accepted. If $T$ approximates $0$, then the formula approximates
  $0$ and the movement is rejected. Hence, $T$ close to $0$ yields a greedy local search.
\item If $d>0$ and if $T$ is an adequate value, then the formula yields a value
  between $0$ and $1$. In this case, we choose a random number uniformly
  from $[0,1]$. If this number is smaller than the value~$e^{-d/T}$ the transition is accepted, otherwise the transition is rejected.
\end{enumerate}

An efficient implementation requires some tweaking, particularly, it is better to use an efficient strategy to select rows~$i$ and $j$. There are several possible approaches:

\begin{enumerate}
\item Choosing $i$ and $j$ uniformly at random, but distinct;
\item Analyzing all pairs $i$ and $j$ and selecting the best existing transition;
\item Using dirty flags, in each row, to speed-up the search of good existing transitions.
\end{enumerate}

\noindent
The following paragraphs provide more details for the previously mentioned approaches.

\subsubsection{Random Choice}
This is the simplest approach and consists of choosing $i$ and $j$
uniformly at random, but distinct. We use this approach if there are no candidates for transitions which reduce $E(H^\prime)$.

\subsubsection{Selecting Best Transitions}
An alternative to the previous approach consists of analyzing
all pairs $i$ and $j$ and selecting the pair leading to the best
existing transition. This requires $O(n m^2)$ time, where $n$ is the number of columns in $H'$ and $m$ is the number of rows. This effort has significant impact in the performance of the resulting algorithm, the time bound is excessive for the resulting gain. Therefore our algorithm never uses this selection procedure. Instead, we propose an heuristic to obtain similar results in at most $O(n m)$ time per analysis. Moreover, our approach only spends this time if there is a chance of reducing the number of ones in $H^\prime$. Most of the time, it certifies that no such move exists in $O(1)$ time.

\subsubsection{Assigning Flags to Rows} Our strategy is to assign a
dirty flag to each row. If this flag is set, then the row is considered dirty, otherwise it is considered clean. If $i$ and $j$ are clean rows, then neither adding row $i$ to row $j$
nor adding row $j$ to row $i$ produces a row with fewer ones. The smaller the number of ones in a matrix, the more rows become clean. Recall, for example, the matrix $H^\prime$ obtained from $H$ in Equation~\eqref{eq:1}. We can safely
consider all rows of $H^\prime$ as clean, because adding any two rows
results in a row which contains more ones. When a given row $i$ is flagged as dirty, we can test this status by adding $i$ to every other row in
$H^\prime$. Note that this is only a testing
procedure and, thus, we do not alter $H^\prime$. If there exists a row~$j$ such that adding $i$ to $j$ reduces
the number of ones, then the pair of rows $i$ and $j$ is sent to the simulated annealing decision process, which, as discussed above, accepts this transition. The dirty flag of $i$ is kept by the simulating annealing procedure and, moreover, row~$j$ is also flagged as dirty. Note that only the selection procedure is able to assign clean flags. A dirty flag is assigned by the simulated annealing process, to row $j$, whenever row $i$ is added to row $j$, no matter what the original flags of $i$ and $j$ are. Also in this case the flag of $i$ is kept.

Row~$i$ can only be flagged as clean if there is no row~$j$ such that adding $i$ to $j$ produces a row with less ones. In this case this selection procedure flags $i$ as clean. This process requires at most $O(nm)$ time, as mentioned before.

Let us now just highlight an important nuance. It is not sufficient to add $i$ to $j$ and verify if this
reduces the number of ones compared to $j$. We must also verify if it reduces the number of ones compared to $i$. If both decrease the number of ones, we choose the transition which results in the sparser matrix. If the number of ones never decreases this selection procedure flags row~$i$ as clean. Moreover, the rows $j$ to test are not considered in order, i.e., from $1$ to $m$. This avoids a bias towards the first rows. Instead this procedure generates a random uniform permutation of the numbers $1$ to $m$.

\subsubsection{Additional Considerations}
In the initialization, all rows are flagged as dirty since none of them was verified. As the algorithm evolves, the number of dirty rows decreases quickly to $0$ and, in fact, most of the time the number of dirty rows is $0$. In this case, no analysis is performed and we select $i$ and $j$ uniformly at random, as explained before. This means that most of the time, we avoid the $O(nm)$ time operation and, instead, spend  $O(1)$ time only,
albeit we also do not obtain decreasing transitions. Still, this process is valuable since when the annealing algorithm escapes a local minima
it quickly moves downhill to another minima. In some cases, we do pay the $O(nm)$ time cost but row $i$ gets flagged as clean. In these cases, we also do not find a row $j$ which decreases the number of ones and again choose $i$ and $j$ uniformly at random.

\subsection{Choosing the Temperature}
\label{sec: ChooseTemperature}
Another important issue in simulated annealing is the selection of the value of the temperature degrees~$T$. This value is not constant during the execution of the algorithm, but it is kept for around 100 iterations. After these iterations, the temperature is updated using a geometric rule, i.e., we change $T$ to $\alpha T$ with $\alpha <1$. Calculating the value of $\alpha$ is straightforward once we decide the initial temperature $T_0$, the final temperature $F$, as well as the number of steps~$s$ desired to transform
$T_0$ into $F$. Hence, the only issue consists in determining the numerical values of $T_0$ and $F$. This is challenging without further insight into the chosen parameters. We provide an intuitive approach to this choice, which is actually very robust for different problems.
\section{Algorithmic Aspects} 
\label{sec:details}
In this section, we review the simulated annealing algorithm and present its
application to the problem of determining the sparsest parity-check matrix. We
also discuss strategies for selecting the temperature and cooling rate. We
finish this section by showing several experimental results of our
approach and surveying the state of the art.
\subsection{Simulated annealing for determining the sparsest PCM}
\label{sec:simulated-annealing}
The simulated annealing algorithm is stated in Algorithm~\ref{algo:sim}. The
current state of the algorithm is represented by the matrix~$H'$ which
starts off equal to $H$. The initial temperature is set to $T_0$ and the
final temperature is set to $F$. When the temperature reaches $F$, the \texttt{while}-loop is executed one last time. Each time this loop is
executed, the \texttt{for}-loop is also executed. The value of
\texttt{Iter} is fixed to 100 in all our tests. We kept it at a low value
within the recommended range. Recall that, for all these 100~values of $k$,
the algorithm iterates at a constant temperature. The temperature is
decreased in line~\ref{line:Td}. To control how long the algorithm
runs, we choose a proper value for $\alpha$. When we want that the external
\texttt{while}-loop is executed $s+1$ (assuming $T_0< F$) times, we set
$\alpha = \sqrt[s]{F/T_0}$.

\begin{algorithm}[htb]
\BlankLine
\SetKwInOut{Input}{input}
\SetKwInOut{Output}{output}

\Input{$\langle H$, $T_0$, $F$, $s\rangle$ \mbox{\{ Matrix $H$ is in alist
    format. \} }}
  \Output{$\langle H^\prime \rangle$}
  \BlankLine
  $H' \leftarrow H$ \;
  $T \leftarrow T_0$ \;
  $\alpha \leftarrow \sqrt[s]{F/T_0}$ \;
  \While{$T \leqslant F$}{
    \For{$k=1$ to \texttt{Iter}}{\label{line:for}
      $(i,j,d) \leftarrow $ \texttt{Analyze($H^\prime$)}
      \label{line:analysis} \;
      \If{{\normalfont \texttt{Random}$(0,1)$} $\leqslant e^{-d/T}$}{ \label{line:if}
        $H^\prime[j] \leftarrow H^\prime[i] + H^\prime[j]$ \; \label{line:sum}
      }
      $k \leftarrow k+1$ \;
    }
    $T \leftarrow \alpha T$
    \label{line:Td} \;
  }
  \Return{$H^\prime$} \;
  \BlankLine
  \BlankLine
\caption{Pseudo code for simulated annealing algorithm.}
\label{algo:sim}
\end{algorithm}

\noindent
The algorithm comprises two functions.

\begin{enumerate}
\item \texttt{Analyze}($H'$) follows the procedure described in
  Section~\ref{sec:my-idea}. If all the rows are marked as clean, then row
  indexes $i$ and $j$ are chosen uniformly at random. We denote the
  respective rows by $H'[i]$ and $H'[j]$. If there is at least one row flagged
  dirty, then $i$ is chosen uniformly among the dirty rows and $j$ scans
  through all the other rows to find one which decreases the number of
  ones. If such a row is found, then $j$ becomes the index of that row.
  In fact, $i$ and $j$ might need to be swapped to maximize the reduction in the
  number of ones. If no such row is found, then the flag of $i$ is set to clean and new $i$ and $j$ values are chosen
  uniformly at random and passed as the output of \texttt{Analyze}. The value~$d$ in line~\ref{line:analysis} represents
  the variation in the number of ones the current transformation will
  imply. If $d$ is negative, the number of ones decreases; if $d$ is
  positive, it increases. The $+$ symbol in line~\ref{line:sum} represents
  the row addition in $\mathbb{Z}_2$.
\item The function \texttt{Random}(0,1) returns a random number in $[0,1]$,
  chosen uniformly at random. To obtain this value and test
  the condition in line~\ref{line:if}, we rewrite the condition. The goal is
  to avoid the loss of precision that results from division and
  exponentiation, and to guarantee sound random numbers. Instead, we test
  the following condition
    \begin{equation*}
        d \leqslant T (30 \ln 2 - \ln R).
    \end{equation*}
    In this condition, $R$ is an integer chosen uniformly at random from $1$
    to $2^{30}$, using the \texttt{arc4random\_uniform} function of the BSD
    \texttt{stdlib}.
\end{enumerate}

It remains to discuss the selection of the temperature parameters $T_0$ and $F$. Like in several parametric methods,
determining these values is largely an experimental procedure, which
depends heavily on the application at hand. This is obviously true for our
application. However, we wish to convey some insight into the choice of
these parameters. Let us recall the acceptance condition in
line~\ref{line:if} of Algorithm~\ref{algo:sim}. The following equation
captures this condition, using $p$ to denote the random number,
\begin{equation}
  p \leqslant e^{-d/T}.
\end{equation}
We may consider the extreme cases, when the inequality becomes an equality,
and we rewrite the condition to obtain $T$. The resulting equation is
\begin{equation}
  T = -d/\ln p.
\label{eq:2}
\end{equation}

This means that a temperature can be defined by specifying $d$ and $p$. We
prefer to specify these parameters as they lead to a more intuitive notion
of temperature. For example, for the matrix $H$ in Equation~\eqref{eq:1}
in Section~\ref{sec:introduction}, we can decide to accept an increase of two
one-entries, i.e., $d=2$, in $4\%$ of the transitions considered, this means
$p = 0.04$. Hence, we obtain a numerical value for $T$ of approximately
$0.62133$. Note that a temperature applies to all the tests of
line~\ref{line:if}, so we might inquire what is the probability that this
temperature accepts a value of $d=4$. Using standard calculus, it turns out
that the resulting probability is $0.04^2=0.0016$. In general, if we define
the values $d_1$ and $p_1$, then the probability $p_2$ for a delta~$d_2$ is given
as $p_2 = p_1^{d_2/d_1}$. Hence, for $d_2=1$, we obtain a probability of
$20\%$ with $p_2 = \sqrt{0.04}=0.2$. The following plot illustrates the
resulting curve.
\begin{figure}[htb]
  \centering
  \centering \includegraphics[scale=0.6]{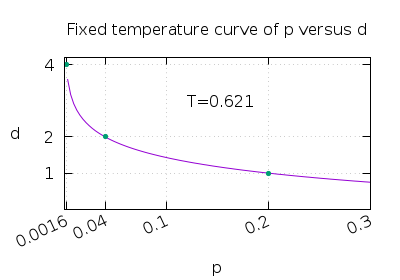}
\end{figure}

Hence, it is possible to specify the same temperature by inputting any one
of the three points indicated above. In general, defining a temperature in this
way is more intuitive than the single numerical value. In our prototype,
temperatures are specified by providing parameters $p$ and $f$. 
The value $d$ is then obtained as $d = f \times N$, where $N$ is the number of
columns of the corresponding matrix.  
We keep the value~$f$ constant at $0.01$, meaning that uphill movements that
increase the number of ones by $d$ are only accepted $1\%$ of the time.
Hence, the input parameters are fairly
intuitive. Naturally the parameters are chosen by trial and error, still
this approach gives some reasonable initial values.

Another important aspect of the simulated annealing algorithm is the
cooling schedule of temperature degrees, which is affected by the final
temperature~$F$. Again, we specify $F$ by choosing $d$ and
$p$. In this case, it is sensible to maintain one of the parameters
constant. For example, we could choose $d=1$ and $p=0.04$, meaning that
from $T_0$ to $F$ we maintain the $4\%$ uphill probability, but reduce the
value of $d$ from $2$ to $1$.

An example of how this approach can simplify the temperature definition is shown in Table~\ref{tab:parameters}, where the selection of hot and cold values for $f$ and $p$ is fairly straightforward, but the resulting temperatures $T_0$ and $F$ are fairly peculiar. Note that, without the insight we have just described, the initial order of magnitude for $T_0$ and amplitude $F-T_0$ are mysteries that need to be solved by trial and error. Let us now proceed to the experimental evaluation.

\subsection{Experimental results and discussion}
\label{sec:experimental-results}
In this section, we describe the experimental setup used to test our
algorithm. We used several PCMs in alist
format\footnote{\url{http://www.inference.org.uk/mackay/codes/alist.html}}. The
matrices are obtained from a channel code
database\footnote{\url{https://www.uni-kl.de/channel-codes/channel-codes-database/}}.
Our prototype is available at \url{https://github.com/LuisRusso-INESC-ID/SPCM}.

\subsubsection{The Design of the Experiments}
\label{sec:design}
We selected reasonable initial temperature~$T_0$ and final temperature~$F$. We also executed our algorithm with extremely low temperature settings. At these
temperatures, the simulated annealing heuristic reduces to a greedy local
search procedure, which always reduces the number of ones and never accepts
any transitions which increase this value.

The result of these executions is shown in plots of time versus the number
of ones in the underlying matrix. Moreover, because the algorithm is
probabilistic, the results vary. Therefore, we present the
results of several repetitions of the algorithm. The local search algorithm
is repeated 32~times and the simulated annealing algorithm is repeated 128~times. The number of times that the \texttt{while}-loop in
Algorithm~\ref{algo:sim} is executed, can be very big, i.e., $s$ can be as high
as several millions. Therefore, we sample data points from some of those
executions.

To execute these tests, we used a dedicated server and executed the tests in
parallel, one per core, but without using hyper-threading. Therefore, we
executed 32 parallel tests at a time. Note that this kind of hardware is
necessary only because we want to study the performance of the proposed
algorithm. For the goal of obtaining a sparse PCM of a
given code, it is possible to use commodity hardware.

The server contained an Intel(R) Xeon(R) CPU E7 4830 running at 2.13GHz,
with 32 physical cores, the architecture is 64-bit. The server contains 4
sockets each containing 8 cores. The system reported 4255.86 BogoMIPS,
and has the following cache sizes L1d 32KB, L1i 32KN, L2 256KB, L3
24576KB. This means that the matrices considered fitted in cache. The
overall system memory is $251$GB and it has $7$GB of swap, but as we just
pointed out, this memory was not crucial to the algorithm. In fact, some of
the tests were initially performed on an Eee PC with an Intel Atom CPU
N270 running at 1.60GHz and with $1.96$GB of main memory.
\subsubsection{Results}
\label{sec:results}
\begin{figure}[htb]
  \centering
  \centering \includegraphics[scale=0.5]{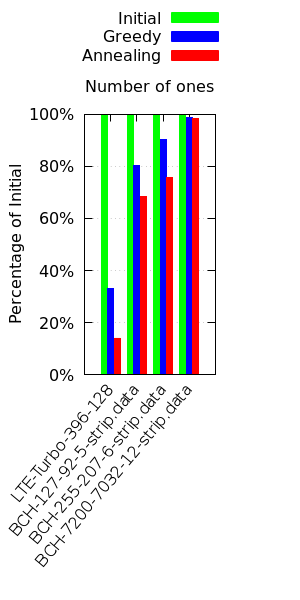}
  \caption{Bar chart illustrating the relative number of ones obtained with
    the greedy and simulated annealing algorithms.}
  \label{fig:bars}
\end{figure}

In Figure~\ref{fig:bars}, we show how the number of ones can be reduced by
our algorithm. The green bars occupy $100$\% of the initial number of
ones. The blue bars indicate the percentage of the number of ones obtained
by the greedy algorithm compared to the initial number of ones. This value
is the one obtained by the best run of the greedy search. The red bars
indicate the percentage of ones obtained by the simulated annealing
algorithm compared to the initial number of ones.

\begin{figure}[htb]
  \centering
  \centering \includegraphics[scale=0.8]{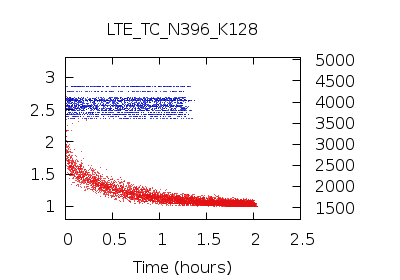}
  \caption{Time ($x$ axis) versus number of ones (right $y$ axis) or factor
  to overall minimum number of ones (left $y$ axis) for LTE\_TC\_N396\_K128.}
  \label{fig:testmatrix2}
\end{figure}

\begin{figure}[htb]
  \centering
  \includegraphics[scale=0.8]{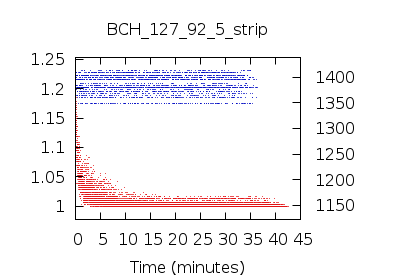}
  \caption{Time ($x$ axis) versus number of ones (right $y$ axis) or factor
  to overall minimum number of ones (left $y$ axis) for BCH\_127\_92\_5\_strip.}
  \label{fig:BCH_127_92_5_strip}
\end{figure}

\begin{figure}[htb]
  \centering
  \includegraphics[scale=0.8]{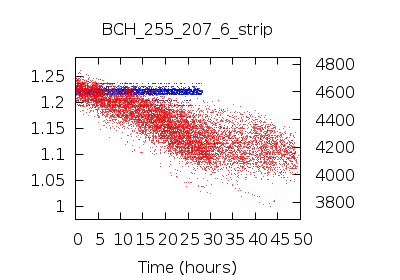}
  \caption{Time ($x$ axis) versus number of ones (right $y$ axis) or factor
  to overall minimum number of ones (left $y$ axis) for BCH\_127\_92\_6\_strip.}
  \label{fig:BCH_255_207_6_strip}
\end{figure}

\begin{figure}[htb]
  \centering
  \includegraphics[scale=0.8]{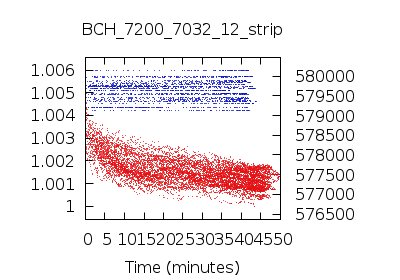}
  \caption{Time ($x$ axis) versus number of ones (right $y$ axis) or factor
  to overall minimum number of ones (left $y$ axis) for BCH\_127\_92\_12\_strip.}
  \label{fig:BCH_7200_7032_12_strip}
\end{figure}

Figures~\ref{fig:testmatrix2},~\ref{fig:BCH_127_92_5_strip},~\ref{fig:BCH_255_207_6_strip},
and~\ref{fig:BCH_7200_7032_12_strip} show the results from the time
experiments. The $x$~axis indicates time, as the algorithm proceeds. The $y$~axis has a double scale, on the right we indicate the actual number of
ones, on the left, we indicate the ratio between the current solution
and the overall minimum attained value. Note that this minimum might be
larger than the number of ones in the code's sparsest parity matrix. The
remaining experimental results are shown in
\ref{sec:experimental-results-1}. The blue points are sampled from
the greedy algorithm and the red points are sampled from the simulated
annealing algorithm. If you have a black and white version of this paper,
then the greedy points are still easy to identify, because they seem to form
constant lines, whereas the simulated annealing points are usually decreasing.

\begin{table}[tb]
  \centering
  \caption{Algorithm parametrization.}

  \label{tab:parameters}
  \scalebox{0.9}{
    \begin{tabular}{l r*{6}{l} r}
      \hline
      & & \multicolumn{3}{c}{Start} & \multicolumn{3}{c}{Finish} & \\
      Code & \multicolumn{1}{c}{$N$} & \multicolumn{1}{c}{$f$} & \multicolumn{1}{c}{$p$} & \multicolumn{1}{c}{$T_0$} & \multicolumn{1}{c}{$f$} & \multicolumn{1}{c}{$p$} & \multicolumn{1}{c}{$F$} & \multicolumn{1}{c}{$S$} \\ \hline
      LTE-TC-N396-K128 & 396 & 0.05 & 0.01 & 4.30 & 0.01 & 0.01 & 0.86 & 5.12E+6 \\
BCH-127-92-5-strip & 127 & 0.05 & 0.01 & 2.77 & 0.01 & 0.01 & 0.28 & 5.12E+6 \\
BCH-255-207-6-strip & 255 & 0.05 & 0.01 & 1.38 & 0.03 & 0.01 & 1.66 & 5.12E+8 \\
BCH-7200-7032-12-strip & 7200 & 0.004 & 0.01 & 6.25 & 0.003 & 0.01 & 4.69 & 1.28E+6 \\
\hline
    \end{tabular}
    }
\end{table}
In Table~\ref{tab:parameters}, we show the parameters that we used for
the simulated annealing algorithm. The number~$N$ indicates the number of
columns of the corresponding matrix. The value of $d$ is obtained as
$f \cdot N$, where $f$ is also given in the table. Likewise, the value~$p$
is also stated in the Table~\ref{tab:parameters}. We use
Equation~\eqref{eq:2} to determine both, the initial $T_0$ and the final
temperature $F$. These values are also shown in the table. The number of iterations~$S$ is given in the last column of
the table. As discussed in Section~\ref{sec:simulated-annealing}, we
keep the value~$f$ constant at $0.01$, meaning that uphill movements that
increase the number of ones by $d$ are only accepted $1\%$ of the time.

\subsubsection{Comments and Discussion}
\label{sec:comments}
The results for the LTE Turbo code with $n=396$ and $k=128$ show the best
illustration for the methods we present. As shown by the bars in
Figure~\ref{fig:bars}, for this particular code, the number of ones in
greedy and simulated annealing algorithms is significantly smaller than the
original number of ones. Moreover, there is even a significant difference
between these two algorithms. The points from the greedy algorithm form
straight lines, showing clearly that this algorithm quickly gets captured
in local minima. The number of ones initially decreases quickly for
both, the greedy and simulated annealing algorithm. It is omitted from the
plots to keep the scale range smaller and make it easier to compare the
performance of the greedy and simulated annealing algorithms. However, for
the LTE Turbo code, it is significant as shown by the bars in
Figure~\ref{fig:bars}.

For the remaining codes, the difference between the greedy and the simulated
annealing is smaller, mainly because the initial number of ones seems to be
closer to the optimal sparsest PCM. Note that the ratios in the scale of
Figures~\ref{fig:BCH_127_92_5_strip},~\ref{fig:BCH_255_207_6_strip},
and~\ref{fig:BCH_7200_7032_12_strip} are very close to $1$ and the blue and
red bars in Figure~\ref{fig:bars} are close to the green bar. The time
plots still show the simulated annealing algorithm improving over time, as
it decreases to reach a factor close to $1$, at which point the searches
seem to stabilize, and possibly the optimal value was obtained. For the
code BCH-255-207-6, the procedure takes much longer, indicating that
for this code better results could be achieved by the simulated annealing
algorithm, given more time. However, since the running time exceeded $50$~hours, the test was stopped. Note that to produce the information in the plot,
we required $5$~times this period.

It is also interesting to note that the difficulty of determining the
sparsest parity-check matrix is intrinsic to the code in consideration and cannot be directly inferred from the number of columns in the code matrix.  The
code BCH-7200-7032-12 has $7200$~columns but seems to stabilize much
quicker, in a matter of minutes, whereas the code BCH-255-207-6 has only
$255$ columns but is much more challenging.

In~\ref{sec:experimental-results-1}, we show further experimental
results. We show several BCH and LTE codes. Our algorithm seems to
consistently and significantly improve the LTE codes. This is a good result
although the improvement ratio seems to degrade as the size of the LTE
codes increases. We believe that this is a parameter problem, as all
these tests are conducted with the same $p$ and $f$ parameters. It seems that this generic scheduling is too cold for the smaller LTE codes, as the cloud of red dots does not seem to contract to a point and, instead, remains wide. When $N$ is between 276
and 612, the cooling seems to be adequate, although for the larger codes
it may benefit from more iterations. For the larger codes, the concavity of the
cloud seems to change, thus indicating that, for these codes, the overall
scheduling is too hot. We plan to experiment with cooler schedules.

As a final consideration, we discuss the practical consequences of reducing
the number of ones in PCMs. As mentioned in Section~\ref{sec:introduction},
our main motivation for sparsifying a PCM is to decrease the time necessary
for ML decoding. A systematic study of ML decoding is beyond the scope of
this paper, and moreover~\cite{DBLP:conf/ict/GensheimerDRKW18} have already
established that ML decoding is indeed faster for sparser matrices. Instead
we focus on code checking, instead of decoding. By checking we mean that
the procedure can only determine whether a codeword $y$ belongs to a
certain code $C$. However when $y$ does not belong to $C$, the checking procedure can
not determine the most likely $y'$ that belongs to $C$ and that got
distorted into $y$.

Restricting our analysis to a checking procedure, instead of a decoder, is motivated
by three reasons. First checking procedures are much simpler than decoders. Second the
number of ones of the PCM has a significant correlation to the performance
of the simple checking procedure we present.  Third, and most importantly, checking is
enough for most of the words. Note that all received words must be checked,
and checking should be enough for most of them, as most of them should be
codewords. Otherwise there will be a significant portion of received words
that will be distorted beyond recovery. This means that checking amounts to
the majority of the time that is necessary to process the received
words. Therefore a reduction in the checking time is guaranteed to
translate into a significant reduction of the overall time, even before the
improvement in the decoding procedure is accounted for.

Now recall that according to definition~\ref{def:code} checking whether a
word $y$ is a code word is a matter computing a matrix multiplication in
$\mathbb{Z}_2$, i.e., $Hy\equiv 0\mod 2$. We will now explain how to
efficiently perform this operation, in a way that depends on the number of
ones in $H$, meaning that the performance of the algorithm benefits from
the fact that $H$ is sparse. First note that we can use xor to compute
addition in $\mathbb{Z}_2$. Second recall the formula for matrix
multiplication, given in the following equation, where the elements of the
resulting vector $v$ are indexed as $v_i$, with $i$ between $1$ and $n-k$,
\begin{equation}
  \label{eq:3}
  v_i = \sum^{n}_{j=1} H_{i,j} y_j.
\end{equation}

The elements where $H_{i,j} = 0$ can simply be removed from the sum. Hence,
in $\mathbb{Z}_2$, this equation can be simplified to
\begin{equation}
  \label{eq:4}
  v_i = \sum_{H_{i,j}=1} y_j.
\end{equation}

This equation is optimised by the fact that $H$ is fixed for several
different words $y$, and depends on the number of ones in $H$. Let us
illustrate this for $y=000000100010111$ and as defined in
Equation~\eqref{eq:1}. The computation in Equation~\ref{eq:3} amounts to
the following calculation, where the $H_{i,j} = 1$ values are highlighted,
\begin{equation*}
\scalebox{0.65}{
$
  \left(
  \begin{array}{c}
    {\colorbox{black!20}{1}}\times 0+0\times 0+{\colorbox{black!20}{1}}\times 0+{\colorbox{black!20}{1}}\times 0+{\colorbox{black!20}{1}}\times 0+0\times 0+0\times
    1+{\colorbox{black!20}{1}}\times 0+{\colorbox{black!20}{1}}\times 0+0\times 0+0\times 1+0\times 0+0\times 1+0\times
    1+0\times 1\\
    0\times 0+{\colorbox{black!20}{1}}\times 0+{\colorbox{black!20}{1}}\times 0+0\times 0+{\colorbox{black!20}{1}}\times 0+0\times 0+0\times
    1+0\times 0+{\colorbox{black!20}{1}}\times 0+0\times 0+0\times 1+0\times 0+0\times 1+0\times
    1+0\times 1\\
    0\times 0+0\times 0+{\colorbox{black!20}{1}}\times 0+{\colorbox{black!20}{1}}\times 0+0\times 0+{\colorbox{black!20}{1}}\times 0+0\times
    1+0\times 0+0\times 0+{\colorbox{black!20}{1}}\times 0+0\times 1+0\times 0+0\times 1+0\times
    1+0\times 1\\
    0\times 0+0\times 0+0\times 0+{\colorbox{black!20}{1}}\times 0+{\colorbox{black!20}{1}}\times 0+0\times 0+{\colorbox{black!20}{1}}\times
    1+0\times 0+0\times 0+0\times 0+{\colorbox{black!20}{1}}\times 1+0\times 0+0\times 1+0\times
    1+0\times 1\\
    0\times 0+0\times 0+0\times 0+0\times 0+{\colorbox{black!20}{1}}\times 0+{\colorbox{black!20}{1}}\times 0+0\times
    1+{\colorbox{black!20}{1}}\times 0+0\times 0+0\times 0+0\times 1+{\colorbox{black!20}{1}}\times 0+0\times 1+0\times
    1+0\times 1\\
    0\times 0+0\times 0+0\times 0+0\times 0+0\times 0+{\colorbox{black!20}{1}}\times 0+{\colorbox{black!20}{1}}\times
    1+0\times 0+{\colorbox{black!20}{1}}\times 0+0\times 0+0\times 1+0\times 0+{\colorbox{black!20}{1}}\times 1+0\times
    1+0\times 1\\
    0\times 0+0\times 0+0\times 0+0\times 0+0\times 0+0\times 0+{\colorbox{black!20}{1}}\times
    1+{\colorbox{black!20}{1}}\times 0+0\times 0+{\colorbox{black!20}{1}}\times 0+0\times 1+0\times 0+0\times 1+{\colorbox{black!20}{1}}\times
    1+0\times 1\\
    0\times 0+0\times 0+0\times 0+0\times 0+0\times 0+0\times 0+0\times
    1+{\colorbox{black!20}{1}}\times 0+{\colorbox{black!20}{1}}\times 0+0\times 0+{\colorbox{black!20}{1}}\times 1+0\times 0+0\times 1+0\times
    1+{\colorbox{black!20}{1}}\times 1\\
  \end{array}
  \right).
$
}
\end{equation*}

On the other hand the computation of the same result according to
Equation~\eqref{eq:4} is illustrated by the following calculation,

\begin{equation*}
\scalebox{0.65}{
$
  \left(
  \begin{array}{c}
     0+ 0+ 0+ 0+ 0+ 0\\
     0+ 0+ 0+ 0\\
     0+ 0+ 0+ 0\\
     0+ 0+
    1+ 1\\
     0+ 0+ 0+ 0\\
     0+
    1+ 0+ 1\\
    1+ 0+ 0+1\\
     0+ 0+ 1+ 1\\
  \end{array}
  \right).
$
}
\end{equation*}

This latter calculation is much smaller than the previous one, thus making
the checking procedure much faster. Hence we use Equation~\eqref{eq:4} to
implement it. A hardware implementation of a checking procedure can definitely
benefit from Equation~\eqref{eq:4}, since every add operation is
implemented with an xor gate and we only need to add the bits $y_j$ for
which $H_{i,j} = 1$. Reducing the number of xor gates reduces both the cost
and the time requirements of the resulting circuit. This application is
clearly important and it will be the focus of further research. For now we
will describe a computer architecture aware checker implementation.

Modern CPUs provide bitwise XOR operations, meaning that the operation is
applied to all the bits in the computer word, i.e., to 64 bits at a time in
contemporary CPUs. The bits in the computer word are processed essentially
in parallel. We explore this parallelism by checking 64 independent
received words at the same time. To clearly establish the relation between
the number of ones in $H$ and the performance of a computation based on
Equation~\eqref{eq:4}, it would be enough to check only one word. However
this would be a considerable waist of performance. Hence we chose to check
64 words at a time. This means that a batch of 64 codewords needs to be
packed for checking, so that all the $y_j$ bits are stored in a single
computer word.

\begin{figure}[htb]
  \centering
  \includegraphics[scale=0.8]{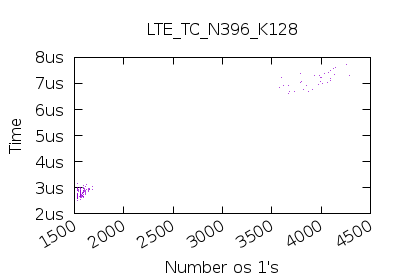}
  \caption{Number of ones ($x$ axis) versus time ($y$ axis).}
  \label{fig:LTE_fastCheck}
\end{figure}

\begin{figure}[htb]
  \centering
  \includegraphics[scale=0.8]{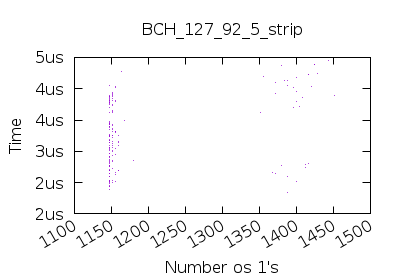}
  \caption{Number of ones ($x$ axis) versus time ($y$ axis).}
  \label{fig:BCH_fastCheck}
\end{figure}

Our test consisted in generating random bit words and checking with several
PCMs for the same code, in particular PCMs obtained with greedy search and
with simulated annealing. The results are shown in the plots in
Figures~\ref{fig:LTE_fastCheck} and~\ref{fig:BCH_fastCheck}.


In these plots the $x$ axis corresponds to the number of ones in the
respective PCMs, and the $y$ axis corresponds to the average time to check a
given word. Like before the results for the LTE Turbo code with $n=396$ and
$k=128$ show the best illustration of the methods we present. In this
example the PCMs with most ones are clearly slower than the PCMs with less
ones. Moreover because the number of ones varies significantly, between
1500 and more than 4000, this effect is even more notorious. This case
clearly shows an important improvement that results from obtaining sparse
PCMs of a given code.

On the other hand for the code BCH-127-92-5-strip this effect is not as
notorious as the variance of the checking time is very big.

\subsubsection{Related Work} 
\label{sec:related-work}

In this section, we discuss some related work to give perspective on
our work. Simulated annealing is a probabilistic technique for
approximating the global minimum of a function. The name derives from the
metallurgy technique of heating and controlled cooling of metal. The method
was initially used to approximate the global minimum of a function with
several variables~\citep*{khachaturyan1979statistical,khachaturyan1981thermodynamic}. It was then formulated in the context of optimization
by~\citet*{Kirkpatrick671}. One of its initial---and most well-known---applications was to the travelling salesman
problem~\citep*{vcerny1985thermodynamical}. Since then, it has been applied to a wide
range of applications, as surveyed by~\citet*{KOULAMAS199441}.

Information theory lies at the core of modern computers and communication
technologies~\citep*{shannon1948mathematical} and can be traced back to the
1940s. The two main applications of information theory are data compression
and error correcting codes. The pioneering work on error correcting codes was
made by~\citet*{6772729} with the introduction of the Hamming(7,4) code. This
code is also a linear code, as the codewords form a linear subspace. The
codewords are 7 bits long. The distance among codewords is $3$ meaning that
is possible to correct errors in a single bit, or detect errors in at most 2
bits.  For a nice introduction to the subject, we refer to the book
of~\citet*{hill1986first}. Several linear codes followed this initial
breakthrough~\citep*{golay1949notes}. The BCH codes that we tested
extensively in this work were discovered independently
by~\citet*{hocquenghem1959codes} and~\citet*{bose1960class}. A personal
description of this time is presented by~\cite{REED200089}, which
naturally describes the history of Reed-Solomon
codes~\citep*{doi:10.1137/0108018}, also a class of linear codes. Turbo
codes are a much more recent discovery and are the first practical codes to
approximate the channel capacity~\citep*{397441}. They are currently in use
in 3G and 4G mobile communication standards and deep space communication, as well as in other
applications where it is necessary to achieve reliable information
transmission over bandwidth or latency constrained channels.

The main advantage of linear codes is that, given their parity-check
matrix, it is straightforward to check if a given word is a codeword or
not. \citet*{Cancellieri2015} gives an extensive discussion on the relation
between generator matrices and PCMs. Detecting errors is simple as
it uses matrix multiplication only. Note that even matrix multiplication
becomes faster for sparser matrices. Still, the main advantage of sparser
matrices is in error correction, for which maximum likelihood approaches
are
used~\citep*{DBLP:journals/ett/BreitbachBLK98,DBLP:phd/ndltd/Feldman03,DBLP:conf/isit/HelmlingRRS14,
  DBLP:journals/tit/TanatmisRHPKW10,DBLP:journals/ett/VontobelK07,DBLP:journals/tit/ZhangS12}. Other
applications of sparse PCMs (in a row sense) along with a
theoretical analysis of lower bounds is presented
by~\citet*{DBLP:journals/combinatorica/NaorV08}.

Recently, \citet*{DBLP:conf/ict/GensheimerDRKW18} pointed out that
minimizing the number of ones in the PCM reduces the run time of
ML~decoding. The authors presented a method of obtaining sparse
PCMs based on integer programming, which can be solved with the
Gurobi or CPLEX solvers. Here, we partially reproduce their result tables
for a couple of BCH and LTE codes and show that the simulated annealing algorithm
obtains comparable final results. Table~\ref{tab:comparison} shows these
results, where the time reflects the amount of time each algorithm was allowed to run. The best result was often found earlier. The IP refers to the algorithm
by~\citeauthor*{DBLP:conf/ict/GensheimerDRKW18}, the bound is a lower bound
on the number of ones. We only computed the lower bound for the smaller BCH codes. As expected, the greedy algorithm performs very well,
but does not always obtain the best value. In most cases, the number of ones
matches the lower bound and are therefore known to be optimal. In the case
of the code BCH-63-36, the lower bound is smaller than the results obtained
by the simulated annealing and the IP algorithm, with both obtaining the
same result of $384$. The results for LTE codes were obtained using the
same simulated annealing parameterization as for BCH codes. Even though the results obtained with simulated annealing (in much less time) are comparable to those obtained by IP, a more tailored parameterization would allow to attain even better results.
\begin{table}[htb]
    \centering
  \caption{Comparison of the number of ones obtained by lower bound, IP,
    the greedy algorithm and the simulated annealing.
    }
\label{tab:comparison}
    \begin{tabular}{ l c c c c c c c }
      \hline
      \multirow{2}{*}{Code} & bound & \multicolumn{2}{c}{IP}  & \multicolumn{2}{c}{Greedy} & \multicolumn{2}{c}{Annealing} \\
                &  \#1s     & \#1s  &  time (s) & \#1s  & time (s) & \#1s & time (s) \\ \hline
      BCH-63-30 &   396     & 396   &  990      & 406   &  191     & 396  &  210 \\  
      BCH-63-36 &   378     & 384   &  810      & 402   &  216     & 384  &  212 \\  
      BCH-63-39 &   336     & 336   &  720      & 344   &  211     & 336  &  221 \\  
      BCH-63-45 &   288     & 288   &  540      & 288   &  301     & 288  &  287 \\  
      BCH-63-51 &   288     & 288   &  360      & 288   &  299     & 288  &  366 \\  
      BCH-63-57 &   192     & 192   &  180      & 192   &  539     & 192  &  630 \\  
      LTE-132-40  & --      & 472   &  82800    & 743   &  5.76    & 562  & 7.71  \\ 
      LTE-156-48  & --      & 568   &  97200    & 940   &  5.83    & 662  & 10.8  \\ 
      LTE-180-56  & --      & 663   &  37200    & 1180  &  6.59    & 776  & 13.8  \\ 
      LTE-204-64  & --      & 760   &  42000    & 1401  &  6.34    & 865  & 18.9  \\ 
      LTE-396-128 & --      & 1594  &  16080    & 3543  &  10.4    & 2030 & 127  \\  
      LTE-780-256 & --      & 3377  &  31440    & 10933 &  20.9    & 5564 & 808  \\  
      \hline
    \end{tabular}
\end{table}

A recent survey on the applications of sparse binary matrices was given
by~\citet*{spm}. The authors start by discussing how to represent sparse
matrices to reduce space requirements and then focus on binary matrices and
their applications in clustering, web graph computations, web link analysis
and binary factor analysis~\citet*{Keprt_2014}.

As a final note we point out the significant impact of the work on graph
sparsification~\citep*{Keprt_2014,benczur1996approximating,Spielman,Spielman_2013,Koutis_2014}
which reduces a graph to a smaller graph that contains many of the same
properties, but that requires much less space. This transformation is
lossy in the sense that it may be impossible to recover the original graph
from the sparse graph. This line of research is a good example of the
advantages that can be obtained with sparser binary matrices, in this case
graph adjacency matrices.

\section{Conclusion and Future work} 
\label{sec:concl-furth-work}
In this paper, we considered the problem of finding the sparsest parity-check matrix (PCM) for a given linear error correcting code. We proposed an
algorithm which modifies this matrix by adding one row to another. We
flagged rows as clean or dirty as a way to speed-up the choice of which rows to alter. For codes
that contained excessive density, this process turns out to be effective.  However,
if this process is applied in a greedy fashion, i.e., never considering
movements which make the underlying matrix denser, the resulting PCM may
still be significantly far from the global optimum. Therefore, we studied a
simulated annealing approach, which yields very good results.

In general, our experimental results indicate that the simulated annealing
algorithm is very likely to achieve a global minimum, given a reasonable
amount of time, with reasonable hardware requirements. Most of the codes
tested seemed to stabilize within a couple of hours with one notable exception
being the BCH-255-207-6 code.

We also proposed a simple way to choose the temperature parameters for the
simulated annealing algorithm, by specifying with what probability $p$ are
we willing to accept an uphill transition that impacts our goal by a value
of $d$. We also explained how this probability $p$ changes for different
values of $d$.

We are currently working on checking the performance of our algorithm in
other error correcting codes and on fine tuning the corresponding ideal
temperatures and cooling schedules. The results of applying simulated
annealing to error decoding were very positive. Moreover, Turbo codes are a
particularly relevant class of codes, see Section~\ref{sec:related-work}
and \ref{sec:experimental-results-1}. In the future, we plan to
investigate more applications of this technique. In particular, we want to apply this technique to the decoding process, thus presenting a
possible alternative to ML decoding.

\section*{Acknowledgements}
  The work reported in this article was supported by national funds through
  Funda\c{c}\~ao para a Ci\^encia e Tecnologia (FCT) with reference
  UID/CEC/50021/2019. This work was funded by European Union's Horizon 2020
  research and innovation programme under the Marie Sk{\l}odowska-Curie
  Actions grant agreement No 690941. This work was supported by DAAD-CRUP
  Luso-German bilateral cooperation under the 2017-2018 research project
  MONO-EMC (Multi-Objective Network Optimization for Engineering and
  Management Support). This work was supported by the DFG (project-ID: RU
  1524/2-3). Jos{\'e} Rui Figueira also acknowledges the support from the
  FCT grant SFRH/BSAB/139892/2018 under POCH Program.

\bibliographystyle{apalike}      
\bibliography{papers}   

\begin{thebibliography}{}

\bibitem[Bencz{\'u}r and Karger, 1996]{benczur1996approximating}
Bencz{\'u}r, A.~A. and Karger, D.~R. (1996).
\newblock {Approximating st Minimum Cuts in $\tilde{O}(n^2)$ Time.}
\newblock In {\em STOC}, volume~96, pages 47--55. Citeseer.

\bibitem[Berlekamp et~al., 1978]{berlekamp}
Berlekamp, E., McEliece, R., and van Tilborg, H. (1978).
\newblock On the inherent intractability of certain coding problems (corresp.).
\newblock {\em IEEE Transactions on Information Theory}, 24(3):384--386.

\bibitem[Berrou et~al., 1993]{397441}
Berrou, C., Glavieux, A., and Thitimajshima, P. (1993).
\newblock Near shannon limit error-correcting coding and decoding: Turbo-codes.
  1.
\newblock In {\em Proceedings of ICC '93 - IEEE International Conference on
  Communications}, volume~2, pages 1064--1070 vol.2.

\bibitem[Bose and Ray-Chaudhuri, 1960]{bose1960class}
Bose, R.~C. and Ray-Chaudhuri, D.~K. (1960).
\newblock On a class of error correcting binary group codes.
\newblock {\em Information and control}, 3(1):68--79.

\bibitem[Breitbach et~al., 1998]{DBLP:journals/ett/BreitbachBLK98}
Breitbach, M., Bossert, M., Lucas, R., and Kempter, C. (1998).
\newblock \emph{Letter} soft-decision decoding of linear block codes as
  optimization problem.
\newblock {\em European Transactions on Telecommunications}, 9(3):289--293.

\bibitem[Cancellieri, 2015]{Cancellieri2015}
Cancellieri, G. (2015).
\newblock {\em Parity Check Matrix Approach to Linear Block Codes}, pages
  245--320.
\newblock Springer International Publishing, Cham.

\bibitem[{\v{C}}ern{\`y}, 1985]{vcerny1985thermodynamical}
{\v{C}}ern{\`y}, V. (1985).
\newblock Thermodynamical approach to the traveling salesman problem: An
  efficient simulation algorithm.
\newblock {\em Journal of optimization theory and applications}, 45(1):41--51.

\bibitem[Feldman, 2003]{DBLP:phd/ndltd/Feldman03}
Feldman, J. (2003).
\newblock {\em Decoding error-correcting codes via linear programming}.
\newblock PhD thesis, Massachusetts Institute of Technology, Cambridge, MA,
  {USA}.

\bibitem[Gensheimer et~al., 2018]{DBLP:conf/ict/GensheimerDRKW18}
Gensheimer, F., Dietz, T., Ruzika, S., Kraft, K., and Wehn, N. (2018).
\newblock Improved maximum-likelihood decoding using sparse parity-check
  matrices.
\newblock In {\em 25th International Conference on Telecommunications, {ICT}
  2018, Saint Malo, France, June 26-28, 2018}, pages 236--240. {IEEE}.

\bibitem[Golay, 1949]{golay1949notes}
Golay, M.~J. (1949).
\newblock Notes on digital coding.
\newblock {\em Proc. IEEE}, 37:657.

\bibitem[Hamming, 1950]{6772729}
Hamming, R.~W. (1950).
\newblock Error detecting and error correcting codes.
\newblock {\em The Bell System Technical Journal}, 29(2):147--160.

\bibitem[Helmling et~al., 2014]{DBLP:conf/isit/HelmlingRRS14}
Helmling, M., Rosnes, E., Ruzika, S., and Scholl, S. (2014).
\newblock Efficient maximum-likelihood decoding of linear block codes on binary
  memoryless channels.
\newblock In {\em 2014 {IEEE} International Symposium on Information Theory,
  Honolulu, HI, USA, June 29 - July 4, 2014}, pages 2589--2593. {IEEE}.

\bibitem[Helmling et~al., 2012]{helmling2012}
Helmling, M., Ruzika, S., and Tanatmis, A. (2012).
\newblock Mathematical programming decoding of binary linear codes: Theory and
  algorithms.
\newblock {\em IEEE Transactions on Information Theory}, 58(7):4753--4769.

\bibitem[Hill, 1986]{hill1986first}
Hill, R. (1986).
\newblock {\em A first course in coding theory}.
\newblock Oxford University Press.

\bibitem[Hocquenghem, 1959]{hocquenghem1959codes}
Hocquenghem, A. (1959).
\newblock Codes correcteurs d’erreurs.
\newblock {\em Chiffres}, 2(2):147--56.

\bibitem[Keprt, 2014]{Keprt_2014}
Keprt, A. (2014).
\newblock Binary matrix pseudo-division and its applications.
\newblock In {\em Innovations in Bio-inspired Computing and Applications},
  pages 153--164. Springer International Publishing.

\bibitem[Khachaturyan et~al., 1979]{khachaturyan1979statistical}
Khachaturyan, A., Semenovskaya, S., and Vainstein, B. (1979).
\newblock A statistical-thermodynamic approach to determination of structure
  amplitude phases.
\newblock {\em Sov. Phys. Crystallography}, 24(5):519--524.

\bibitem[Khachaturyan et~al., 1981]{khachaturyan1981thermodynamic}
Khachaturyan, A., Semenovsovskaya, S., and Vainshtein, B. (1981).
\newblock The thermodynamic approach to the structure analysis of crystals.
\newblock {\em Acta Crystallographica Section A: Crystal Physics, Diffraction,
  Theoretical and General Crystallography}, 37(5):742--754.

\bibitem[Kirkpatrick et~al., 1983]{Kirkpatrick671}
Kirkpatrick, S., Gelatt, C.~D., and Vecchi, M.~P. (1983).
\newblock Optimization by simulated annealing.
\newblock {\em Science}, 220(4598):671--680.

\bibitem[Koulamas et~al., 1994]{KOULAMAS199441}
Koulamas, C., Antony, S., and Jaen, R. (1994).
\newblock A survey of simulated annealing applications to operations research
  problems.
\newblock {\em Omega}, 22(1):41 -- 56.

\bibitem[Koutis et~al., 2014]{Koutis_2014}
Koutis, I., Miller, G.~L., and Peng, R. (2014).
\newblock Approaching optimality for solving {SDD} linear systems.
\newblock {\em {SIAM} Journal on Computing}, 43(1):337--354.

\bibitem[Martinovic et~al., 2005]{spm}
Martinovic, J., Dvorsk, J., and Snasel, V. (2005).
\newblock Sparse binary matrices.
\newblock {\em ITAT 2005 - Workshop on Theory and Practice of Information
  Technologies - Applications and Theory, Proceedings}.

\bibitem[Naor and Verstra{\"{e}}te, 2008]{DBLP:journals/combinatorica/NaorV08}
Naor, A. and Verstra{\"{e}}te, J. (2008).
\newblock Parity check matrices and product representations of squares.
\newblock {\em Combinatorica}, 28(2):163--185.

\bibitem[Reed, 2000]{REED200089}
Reed, I. (2000).
\newblock A brief history of the development of error correcting codes.
\newblock {\em Computers \& Mathematics with Applications}, 39(11):89 -- 93.

\bibitem[Reed and Solomon, 1960]{doi:10.1137/0108018}
Reed, I. and Solomon, G. (1960).
\newblock Polynomial codes over certain finite fields.
\newblock {\em Journal of the Society for Industrial and Applied Mathematics},
  8(2):300--304.

\bibitem[Shannon, 1948]{shannon1948mathematical}
Shannon, C.~E. (1948).
\newblock A mathematical theory of communication.
\newblock {\em Bell system technical journal}, 27(3):379--423.

\bibitem[Spielman and Teng, 2013a]{Spielman}
Spielman, D. and Teng, S.-H. (2013a).
\newblock Solving sparse, symmetric, diagonally-dominant linear systems in time
  o(m/sup 1.31/.
\newblock In {\em 44th Annual {IEEE} Symposium on Foundations of Computer
  Science, 2003. Proceedings.} {IEEE} Computer. Soc.

\bibitem[Spielman and Teng, 2013b]{Spielman_2013}
Spielman, D.~A. and Teng, S.-H. (2013b).
\newblock A local clustering algorithm for massive graphs and its application
  to nearly linear time graph partitioning.
\newblock {\em {SIAM} Journal on Computing}, 42(1):1--26.

\bibitem[Tanatmis et~al., 2010]{DBLP:journals/tit/TanatmisRHPKW10}
Tanatmis, A., Ruzika, S., Hamacher, H.~W., Punekar, M., Kienle, F., and Wehn,
  N. (2010).
\newblock A separation algorithm for improved lp-decoding of linear block
  codes.
\newblock {\em {IEEE} Trans. Information Theory}, 56(7):3277--3289.

\bibitem[Vontobel and Koetter, 2007]{DBLP:journals/ett/VontobelK07}
Vontobel, P.~O. and Koetter, R. (2007).
\newblock On low-complexity linear-programming decoding of {LDPC} codes.
\newblock {\em European Transactions on Telecommunications}, 18(5):509--517.

\bibitem[Zhang and Siegel, 2012]{DBLP:journals/tit/ZhangS12}
Zhang, X. and Siegel, P.~H. (2012).
\newblock Adaptive cut generation algorithm for improved linear programming
  decoding of binary linear codes.
\newblock {\em {IEEE} Trans. Information Theory}, 58(10):6581--6594.

\end{thebibliography}

\vfill\newpage

\appendix

\section{Experimental Results}
\label{sec:experimental-results-1}

This appendix provides more experimental results for our algorithm.

This first plot is similar to the plot in Figure~\ref{fig:bars}, but for our full test set. The bars represent percentage of the number of ones in the resulting PCMs. The green bars are always 100\%. The blue bars correspond to the results obtained by the greedy approach and the red bars the results of the Simulated Annealing algorithm. The results are sorted in decreasing values of the greedy algorithm.

\begin{figure}[htb]
  \centering
  \centering \includegraphics{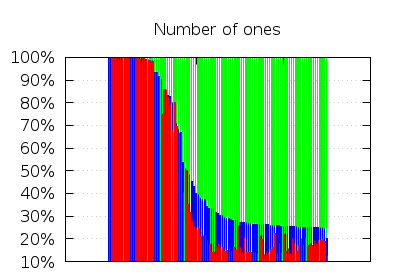}
  \caption{Bar chart illustrating the relative number of ones obtained with
    the greedy and simulated annealing algorithms.}
\end{figure}

\vspace{2cm}

\begin{figure}[htb]
\centering
  \begin{minipage}[b]{.4\textwidth}
    \centering \includegraphics[scale=0.45]{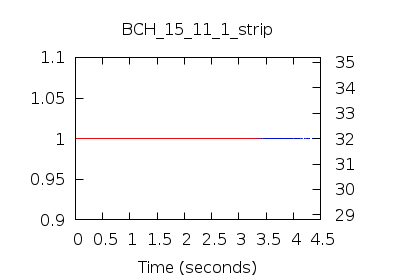}
  \end{minipage}
  \begin{minipage}[b]{.4\textwidth}
    \centering \includegraphics[scale=0.45]{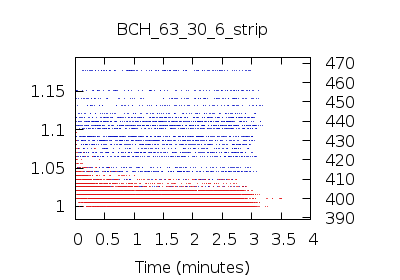}
  \end{minipage}
  \begin{minipage}[b]{.4\textwidth}
    \centering \includegraphics[scale=0.45]{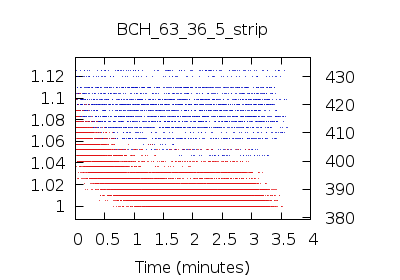}
  \end{minipage}
  \begin{minipage}[b]{.4\textwidth}
    \centering \includegraphics[scale=0.45]{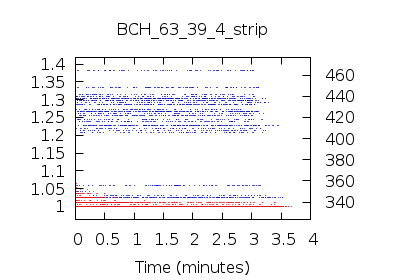}
  \end{minipage}
  \begin{minipage}[b]{.4\textwidth}
    \centering \includegraphics[scale=0.45]{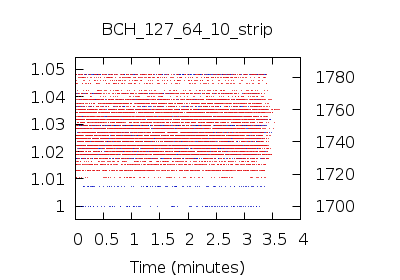}
  \end{minipage}
  \begin{minipage}[b]{.4\textwidth}
    \centering \includegraphics[scale=0.45]{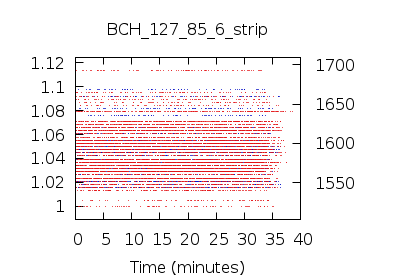}
  \end{minipage}
\end{figure}

\begin{figure}[htb]
   \begin{minipage}[b]{.4\textwidth}
    \centering \includegraphics[scale=0.45]{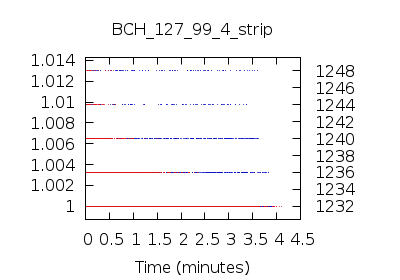}
  \end{minipage}
  \begin{minipage}[b]{.4\textwidth}
    \centering \includegraphics[scale=0.45]{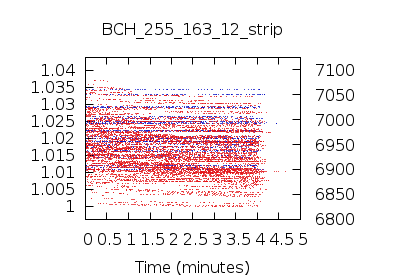}
  \end{minipage}
  \begin{minipage}[b]{.4\textwidth}
    \centering \includegraphics[scale=0.45]{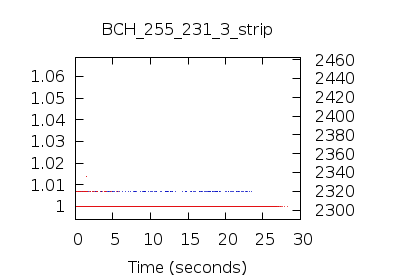}
  \end{minipage}
    \begin{minipage}[b]{.4\textwidth}
    \centering \includegraphics[scale=0.45]{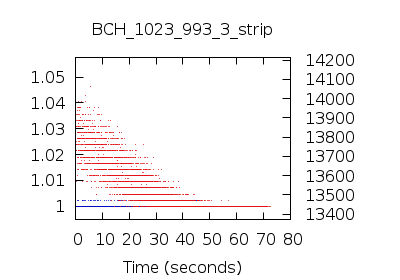}
  \end{minipage}
  \centering
  \begin{minipage}[b]{.4\textwidth}
    \centering \includegraphics[scale=0.45]{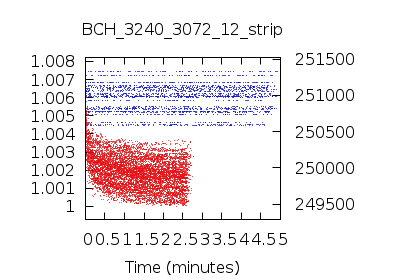}
  \end{minipage}
  \begin{minipage}[b]{.4\textwidth}
    \centering \includegraphics[scale=0.45]{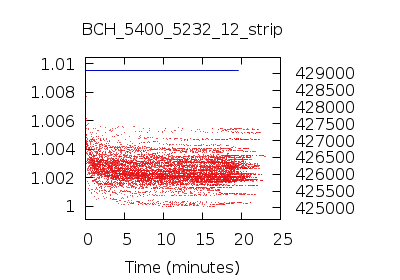}
  \end{minipage}
  \begin{minipage}[b]{.4\textwidth}
    \centering \includegraphics[scale=0.45]{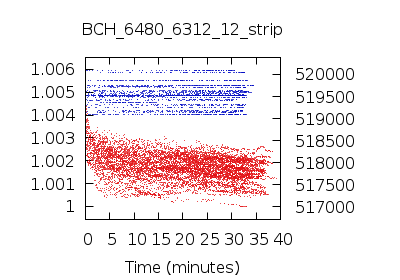}
  \end{minipage}
  \begin{minipage}[b]{.4\textwidth}
    \centering \includegraphics[scale=0.45]{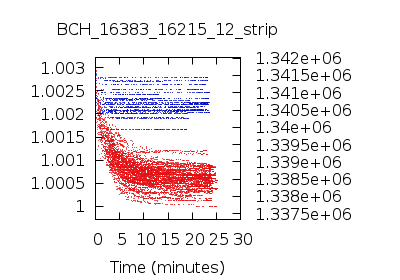}
  \end{minipage}
\end{figure}

\begin{figure}[htb]
\begin{minipage}[b]{.4\textwidth}
  \centering \includegraphics[scale=0.45]{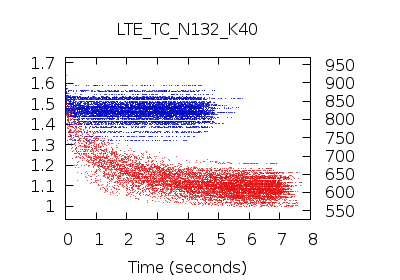}
\end{minipage}
\begin{minipage}[b]{.4\textwidth}
  \centering \includegraphics[scale=0.45]{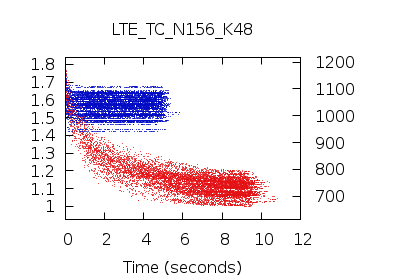}
\end{minipage}
\begin{minipage}[b]{.4\textwidth}
  \centering \includegraphics[scale=0.45]{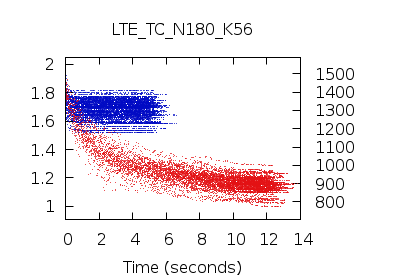}
\end{minipage}
\begin{minipage}[b]{.4\textwidth}
  \centering \includegraphics[scale=0.45]{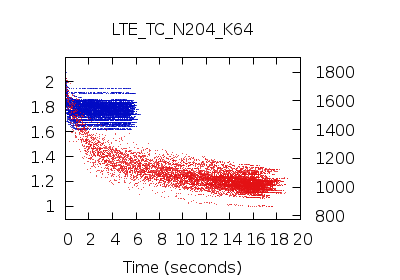}
\end{minipage}
\centering
\begin{minipage}[b]{.4\textwidth}
  \centering \includegraphics[scale=0.45]{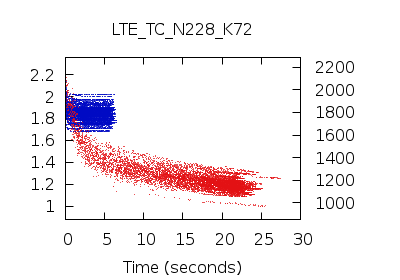}
\end{minipage}
\begin{minipage}[b]{.4\textwidth}
  \centering \includegraphics[scale=0.45]{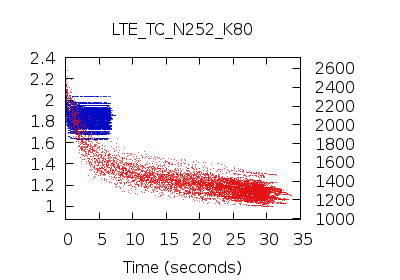}
\end{minipage}
\begin{minipage}[b]{.4\textwidth}
  \centering \includegraphics[scale=0.45]{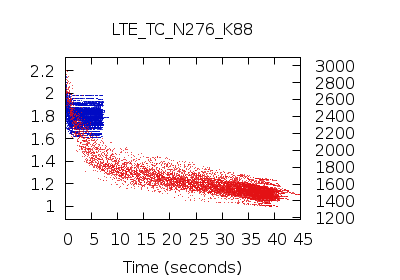}
\end{minipage}
\begin{minipage}[b]{.4\textwidth}
  \centering \includegraphics[scale=0.45]{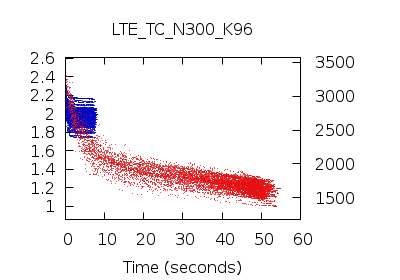}
\end{minipage}
\end{figure}

\begin{figure}[htb]
\begin{minipage}[b]{.4\textwidth}
  \centering \includegraphics[scale=0.45]{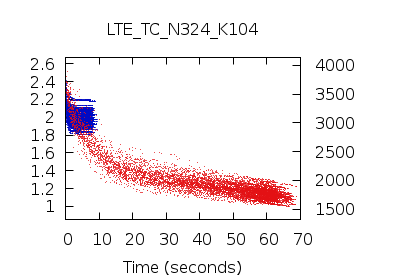}
\end{minipage}
\begin{minipage}[b]{.4\textwidth}
  \centering \includegraphics[scale=0.45]{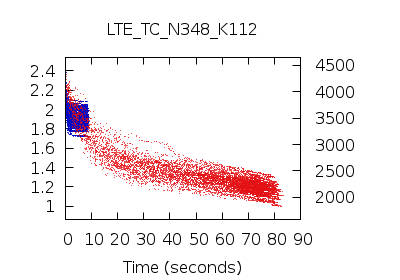}
\end{minipage}
\begin{minipage}[b]{.4\textwidth}
  \centering \includegraphics[scale=0.45]{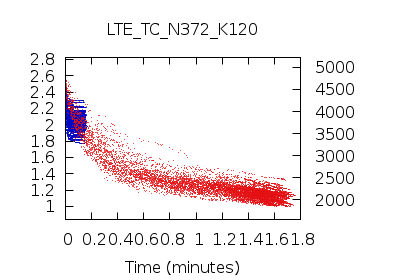}
\end{minipage}
\begin{minipage}[b]{.4\textwidth}
  \centering \includegraphics[scale=0.45]{LTE_TC_N396_K128.png}
\end{minipage}
\centering
\begin{minipage}[b]{.4\textwidth}
  \centering \includegraphics[scale=0.45]{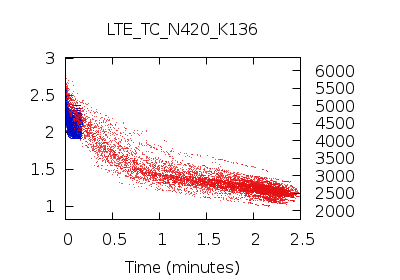}
\end{minipage}
\begin{minipage}[b]{.4\textwidth}
  \centering \includegraphics[scale=0.45]{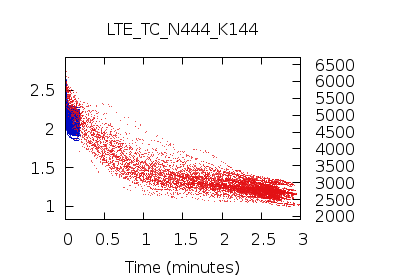}
\end{minipage}
\begin{minipage}[b]{.4\textwidth}
  \centering \includegraphics[scale=0.45]{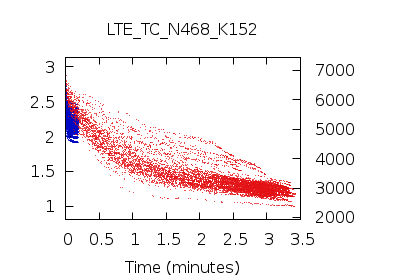}
\end{minipage}
\begin{minipage}[b]{.4\textwidth}
  \centering \includegraphics[scale=0.45]{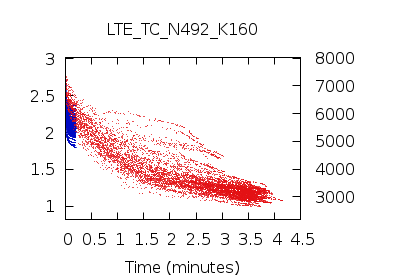}
\end{minipage}
\end{figure}

\begin{figure}[htb]
\begin{minipage}[b]{.4\textwidth}
  \centering \includegraphics[scale=0.45]{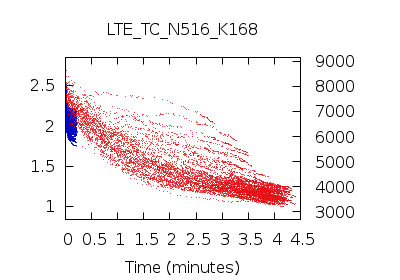}
\end{minipage}
\begin{minipage}[b]{.4\textwidth}
  \centering \includegraphics[scale=0.45]{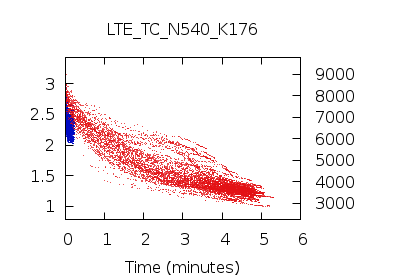}
\end{minipage}
\begin{minipage}[b]{.4\textwidth}
  \centering \includegraphics[scale=0.45]{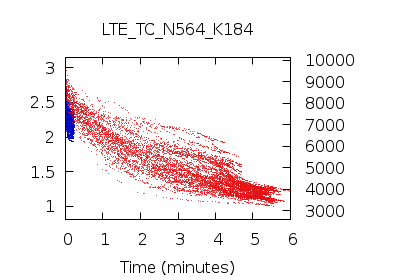}
\end{minipage}
\begin{minipage}[b]{.4\textwidth}
  \centering \includegraphics[scale=0.45]{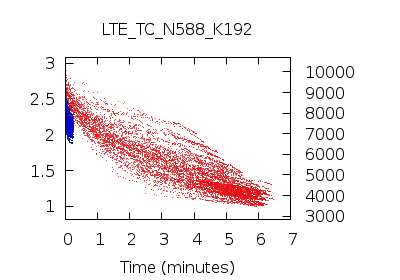}
\end{minipage}
\centering
\begin{minipage}[b]{.4\textwidth}
  \centering \includegraphics[scale=0.45]{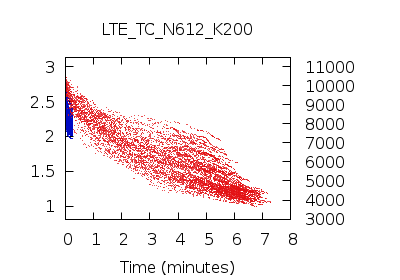}
\end{minipage}
\begin{minipage}[b]{.4\textwidth}
  \centering \includegraphics[scale=0.45]{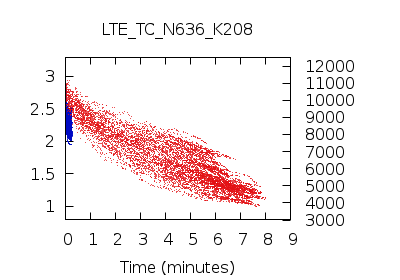}
\end{minipage}
\begin{minipage}[b]{.4\textwidth}
  \centering \includegraphics[scale=0.45]{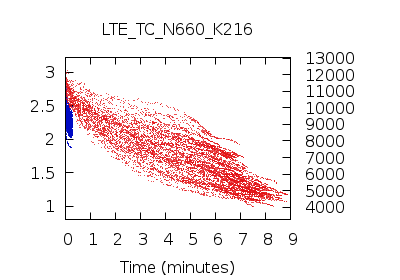}
\end{minipage}
\begin{minipage}[b]{.4\textwidth}
  \centering \includegraphics[scale=0.45]{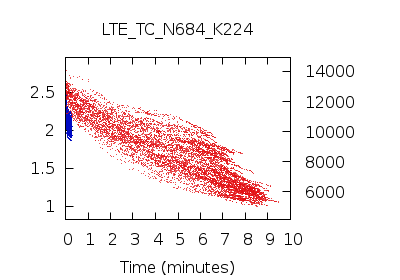}
\end{minipage}
\end{figure}

\begin{figure}[htb]
\begin{minipage}[b]{.4\textwidth}
  \centering \includegraphics[scale=0.45]{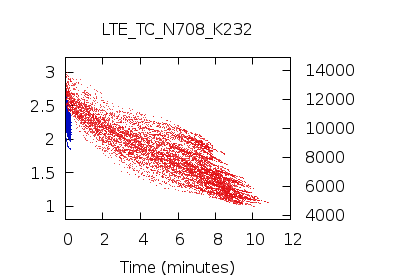}
\end{minipage}
\begin{minipage}[b]{.4\textwidth}
  \centering \includegraphics[scale=0.45]{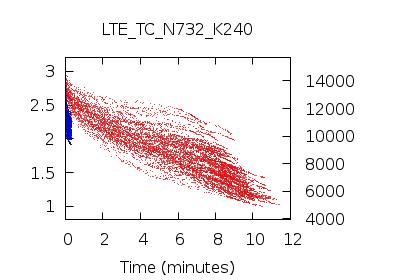}
\end{minipage}
\begin{minipage}[b]{.4\textwidth}
  \centering \includegraphics[scale=0.45]{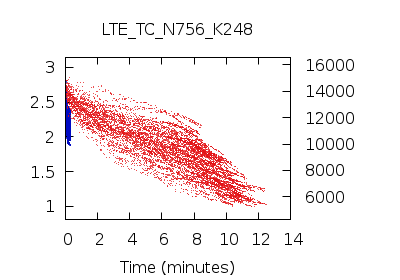}
\end{minipage}
\begin{minipage}[b]{.4\textwidth}
  \centering \includegraphics[scale=0.45]{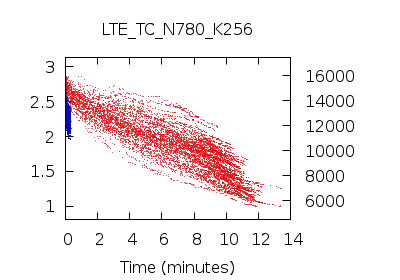}
\end{minipage}
\centering
\begin{minipage}[b]{.4\textwidth}
  \centering \includegraphics[scale=0.45]{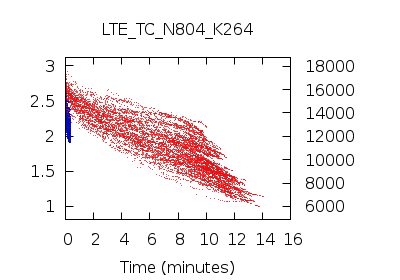}
\end{minipage}
\begin{minipage}[b]{.4\textwidth}
  \centering \includegraphics[scale=0.45]{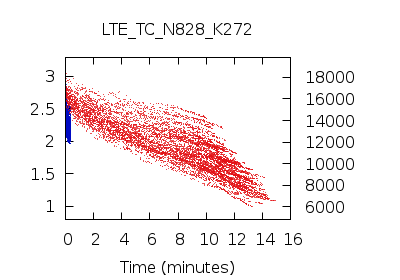}
\end{minipage}
\begin{minipage}[b]{.4\textwidth}
  \centering \includegraphics[scale=0.45]{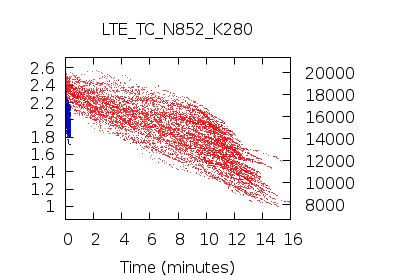}
\end{minipage}
\begin{minipage}[b]{.4\textwidth}
  \centering \includegraphics[scale=0.45]{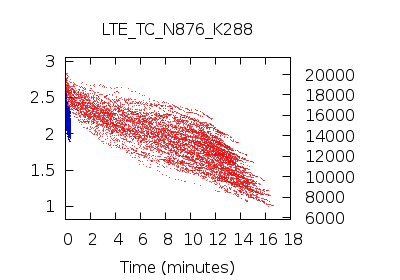}
\end{minipage}
\end{figure}

\begin{figure}[htb]
\begin{minipage}[b]{.4\textwidth}
  \centering \includegraphics[scale=0.45]{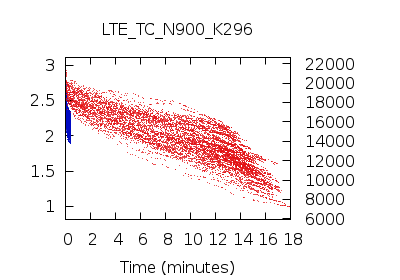}
\end{minipage}
\begin{minipage}[b]{.4\textwidth}
  \centering \includegraphics[scale=0.45]{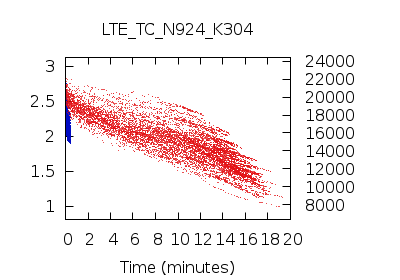}
\end{minipage}
\begin{minipage}[b]{.4\textwidth}
  \centering \includegraphics[scale=0.45]{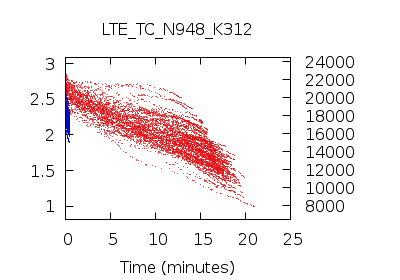}
\end{minipage}
\begin{minipage}[b]{.4\textwidth}
  \centering \includegraphics[scale=0.45]{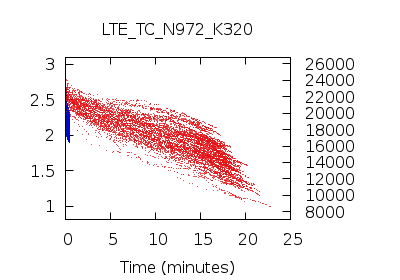}
\end{minipage}
\centering
\begin{minipage}[b]{.4\textwidth}
  \centering \includegraphics[scale=0.45]{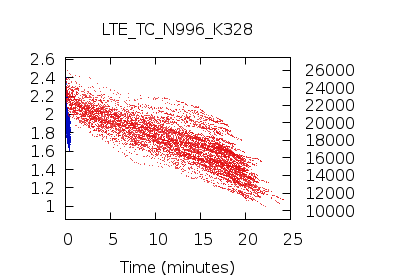}
\end{minipage}
\begin{minipage}[b]{.4\textwidth}
  \centering \includegraphics[scale=0.45]{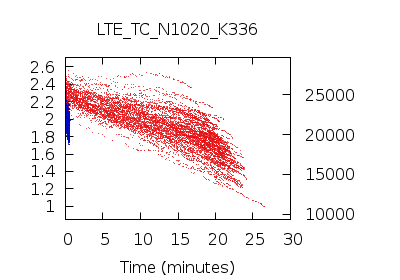}
\end{minipage}
\begin{minipage}[b]{.4\textwidth}
  \centering \includegraphics[scale=0.45]{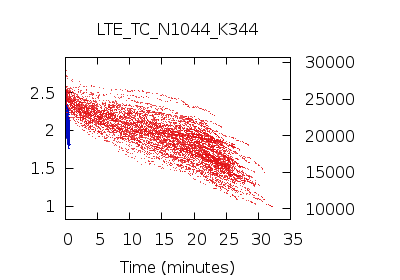}
\end{minipage}
\begin{minipage}[b]{.4\textwidth}
  \centering \includegraphics[scale=0.45]{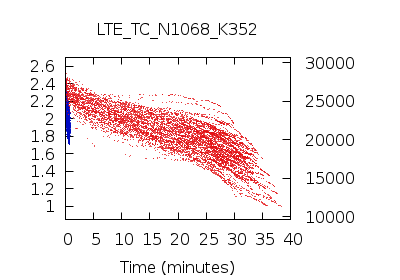}
\end{minipage}
\end{figure}

\begin{figure}[htb]
\begin{minipage}[b]{.4\textwidth}
  \centering \includegraphics[scale=0.45]{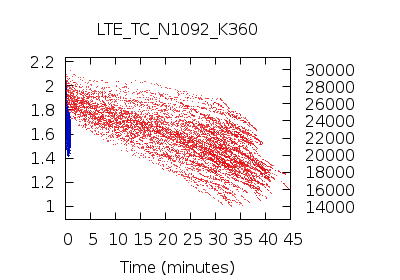}
\end{minipage}
\begin{minipage}[b]{.4\textwidth}
  \centering \includegraphics[scale=0.45]{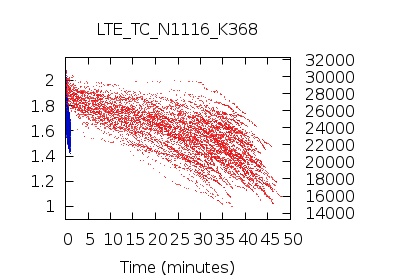}
\end{minipage}
\begin{minipage}[b]{.4\textwidth}
  \centering \includegraphics[scale=0.45]{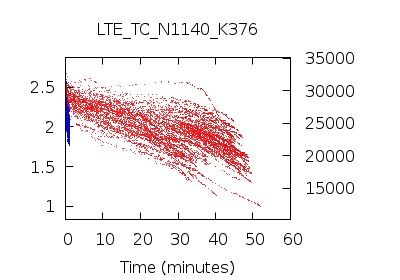}
\end{minipage}
\begin{minipage}[b]{.4\textwidth}
  \centering \includegraphics[scale=0.45]{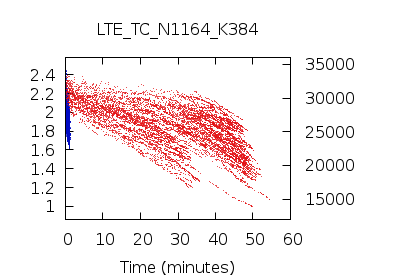}
\end{minipage}
\centering
\begin{minipage}[b]{.4\textwidth}
  \centering \includegraphics[scale=0.45]{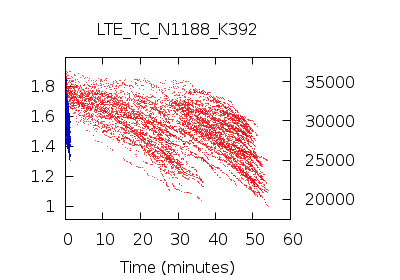}
\end{minipage}
\begin{minipage}[b]{.4\textwidth}
  \centering \includegraphics[scale=0.45]{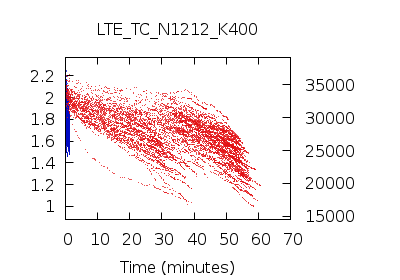}
\end{minipage}
\begin{minipage}[b]{.4\textwidth}
  \centering \includegraphics[scale=0.45]{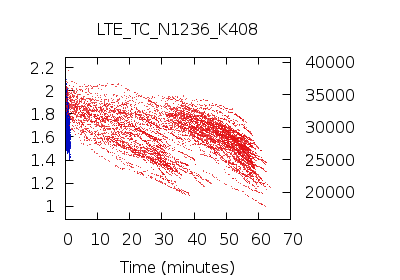}
\end{minipage}
\begin{minipage}[b]{.4\textwidth}
  \centering \includegraphics[scale=0.45]{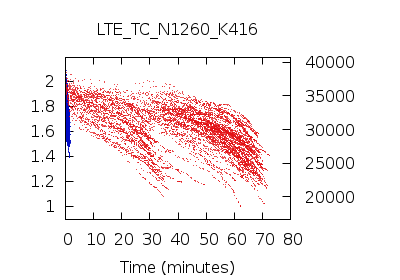}
\end{minipage}
\end{figure}

\begin{figure}[htb]
\begin{minipage}[b]{.4\textwidth}
  \centering \includegraphics[scale=0.45]{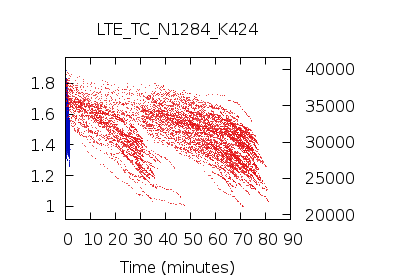}
\end{minipage}
\begin{minipage}[b]{.4\textwidth}
  \centering \includegraphics[scale=0.45]{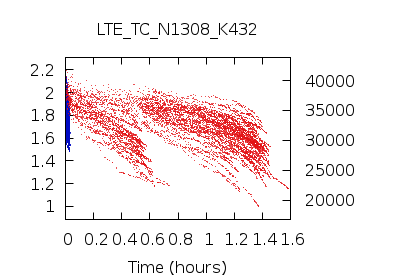}
\end{minipage}
\begin{minipage}[b]{.4\textwidth}
  \centering \includegraphics[scale=0.45]{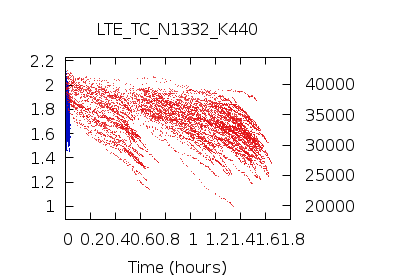}
\end{minipage}
\begin{minipage}[b]{.4\textwidth}
  \centering \includegraphics[scale=0.45]{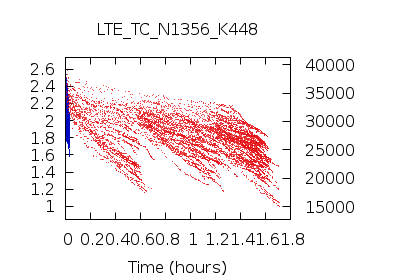}
\end{minipage}
\centering
\begin{minipage}[b]{.4\textwidth}
  \centering \includegraphics[scale=0.45]{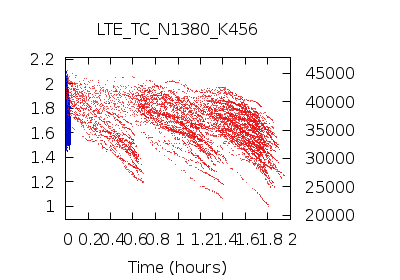}
\end{minipage}
\begin{minipage}[b]{.4\textwidth}
  \centering \includegraphics[scale=0.45]{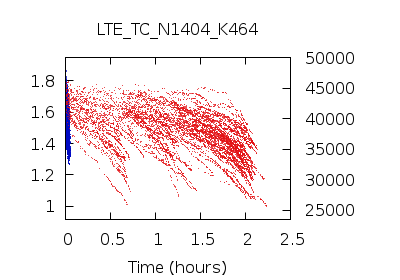}
\end{minipage}
\begin{minipage}[b]{.4\textwidth}
  \centering \includegraphics[scale=0.45]{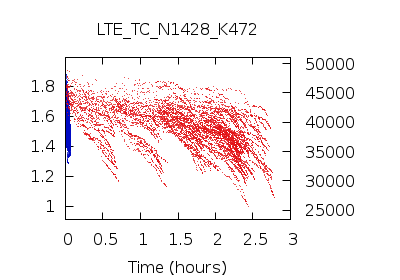}
\end{minipage}
\begin{minipage}[b]{.4\textwidth}
  \centering \includegraphics[scale=0.45]{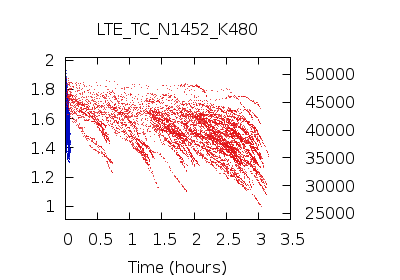}
\end{minipage}
\end{figure}

\begin{figure}[htb]
\begin{minipage}[b]{.4\textwidth}
  \centering \includegraphics[scale=0.45]{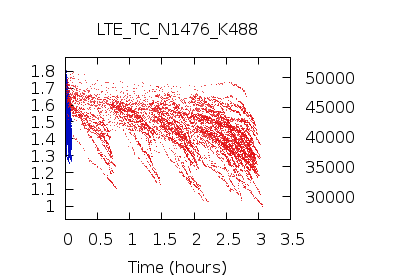}
\end{minipage}
\begin{minipage}[b]{.4\textwidth}
  \centering \includegraphics[scale=0.45]{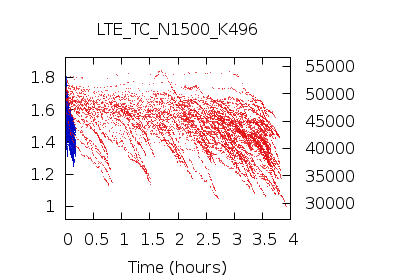}
\end{minipage}
\begin{minipage}[b]{.4\textwidth}
  \centering \includegraphics[scale=0.45]{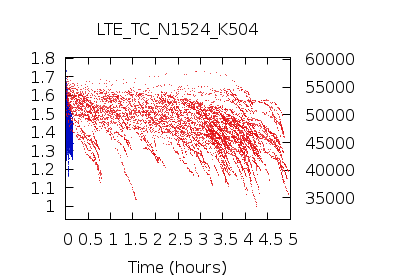}
\end{minipage}
\begin{minipage}[b]{.4\textwidth}
  \centering \includegraphics[scale=0.45]{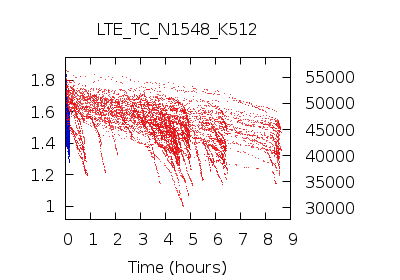}
\end{minipage}
\centering
\begin{minipage}[b]{.4\textwidth}
  \centering \includegraphics[scale=0.45]{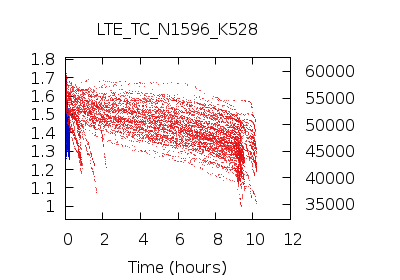}
\end{minipage}
\begin{minipage}[b]{.4\textwidth}
  \centering \includegraphics[scale=0.45]{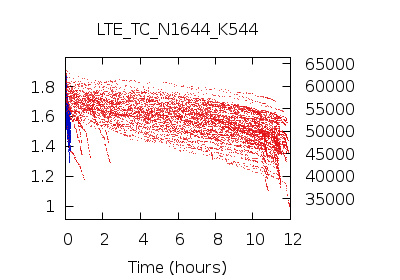}
\end{minipage}
\begin{minipage}[b]{.4\textwidth}
  \centering \includegraphics[scale=0.45]{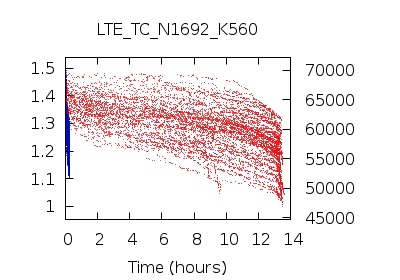}
\end{minipage}
\begin{minipage}[b]{.4\textwidth}
  \centering \includegraphics[scale=0.45]{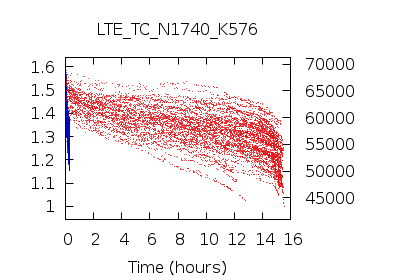}
\end{minipage}
\end{figure}

\end{document}